\documentclass[prb,aps,twocolumn,amsmath,amssymb,floatfix,superscriptaddress]{revtex4}
\usepackage{amsmath}
\usepackage{amsfonts}
\usepackage{amssymb}
\usepackage[dvips]{graphics}
\usepackage{color}
\definecolor{dred}{rgb}{0.75,0,0}
\usepackage{soul}
\usepackage[colorlinks=true, citecolor=blue, urlcolor=blue ]{hyperref}

\date{\today}

\begin{document}
	
\title{Modulation of charge and spin circular currents in a ring-wire hybrid setup} 
  
\author{Joydeep Majhi}


\affiliation{Physics and Applied Mathematics Unit, Indian Statistical Institute, 203, Barrackpore Trunk Road, Kolkata-700 108, India}
	
\author{Santanu K. Maiti}


\affiliation{Physics and Applied Mathematics Unit, Indian Statistical Institute, 203, Barrackpore Trunk Road, Kolkata-700 108, India}

\begin{abstract}
\href{mailto:~~joydeepmjh@gmail.com}{joydeepmjh@gmail.com}  (Joydeep Majhi), \href{mailto:~~santanu.maiti@isical.ac.in}{santanu.maiti@isical.ac.in} (Santanu K. Maiti)\\
\vskip .5cm

We present a comprehensive investigation into the charge and spin circular currents in a mesoscopic hybrid system, with a particular 
focus on the intricate interplay between the Aubry-Andr\'e-Harper (AAH) potential, Aharonov-Bohm (AB) flux, chemical potential ($\mu$), 
and antiferromagnetic (AF) ordering. The proposed quantum system comprises a composite structure of an AF ring coupled to an AAH chain.
Utilizing a tight-binding model and operator method to calculate charge and spin circular currents, we uncover a range of intriguing 
phenomena. An interesting finding is that while the antiferromagnetic ring alone does not exhibit spin channel separation due to the 
symmetry between the up and down spin sub-Hamiltonians, the introduction of a dangling bond or chain can break this symmetry, leading 
to a spin separation effect. The AAH potential in the chain disrupts energy level symmetry, significantly impacting transport behavior.
The charge circular current exhibits periodic oscillations with the AB flux, and its polarity changes with variations in $\mu$. The AAH 
phase affects the charge current through the shifting of energy levels. Furthermore, the spin current displays oscillatory behavior as 
the AAH potential strength changes, with peaks emerging due to interference phenomena caused by disorder-induced localization. 
A possible experimental realization of our proposed quantum setup is also discussed, for the sake of completeness.
This work may provide important insights into the complex physics of charge and spin transport in various hybrid mesoscopic systems, 
offering promising avenues for future research and technological applications.

\end{abstract}

\maketitle

\section{Introduction}

Mesoscopic systems~\cite{dat1,Imry1998,Akkermans2007}, with sizes between individual atoms and bulk materials, have attracted 
substantial attention due to their unique electronic and magnetic properties, including the interesting phenomenon of persistent
current~\cite{Price2020,Yerin2021,Chetcuti2022,Pace2022,Loss1992}. The current, whether charge or spin, flows without external bias 
and arises from quantum interference effect within mesoscopic rings or quantum dots. Studying of charge and spin dependent persistent 
currents is not only fundamental for understanding quantum mechanics in condensed matter but also holds promise for designing advanced 
nanoscale electronic devices~\cite{li2018}, with particular significance in the field of spintronics, aiming to exploit electron spin 
for novel electrical devices~\cite{vzutic2004,zutic2006,pulizzi2012,hirohata2020}.

For proper development of spintronic devices, the manipulation of electron spin and thus the spin current is of the 
utmost importance~\cite{hirohata2020}. For decades, ferromagnetic materials were thought to be the most important functional
elements~\cite{jansen2012,lima2021,tanaka2020,maassen2011}. However, with the successful incorporation of new functional elements,
antiferromagnetic systems, the specific roles of ferromagnetic materials are getting reduced in spintronic
applications~\cite{jungwirth2016,wang2022,baltz2018}. The ability of producing high-speed, low-power spintronics devices from 
antiferromagnetic materials has spawned a new field of study which is referred to as antiferromagnetic
spintronics~\cite{vsmejkal2018,han2023,jungfleisch2018}. 
Unlike ferromagnetic components, antiferromagnets do not have any net magnetization. The zero net magnetization mitigates the 
issue of stray fields interfering with nearby devices~\cite{xiong2022}. We can pack more functional elements in a given space 
without compromising any efficiency. Data processing speeds can also be increased by using antiferromagnetic materials because of their 
faster and more efficient spin dynamics~\cite{tang2018}.

The ability of antiferromagnetism to induce spin-polarized currents in neighboring materials is highly important~\cite{seki2015}. 
The spin of an electron can be transferred to the magnetization of an adjacent layer via a spin-transfer torque, creating a 
spin-polarized current~\cite{slonczewski2007,sankey2008,chureemart2011}. This effect could be used in spintronic devices like 
MRAM and spin valves to store and retrieve information~\cite{mirowski2007,parkin2003}. Many interesting physical properties are seen in
antiferromagnetic materials, such as spin waves for information transmission and magnetization dynamics that can be controlled by applied
fields. Antiferromagnetism is significant in mesoscopic physics and spintronics because of its unusual properties and the possibility that 
they will open the door to new types of novel devices and methods of processing data~\cite{wang2022,baltz2018,vsmejkal2018,han2023}.

The interaction between quantum systems and the influence of their basic factors has always been a fascinating area in
physics~\cite{magarill1996,recher2007,russo2008,majhi2020}. In this context, we explore an interesting area where the behaviors 
of an antiferromagnetic ring and a one-dimensional chain connect. We are motivated by the strong desire to understand how 
symmetry-breaking processes affect electronic transport in quantum systems~\cite{dat1,ren2015,trzeciecki2000}. Achieving spin-resolved 
channels in antiferromagnetic material is a difficult task, but it is not impossible. And in this study, \textit{we propose a simple 
and elegant method for removing the degeneracy between up and down spin states, allowing antiferromagnetic systems to support a net 
spin current that can be manipulated by tuning AB flux and other system parameters.} This is the primary motivation of our work.

\begin{figure}[ht]
{\centering \resizebox*{8.2cm}{7.4cm}{\includegraphics{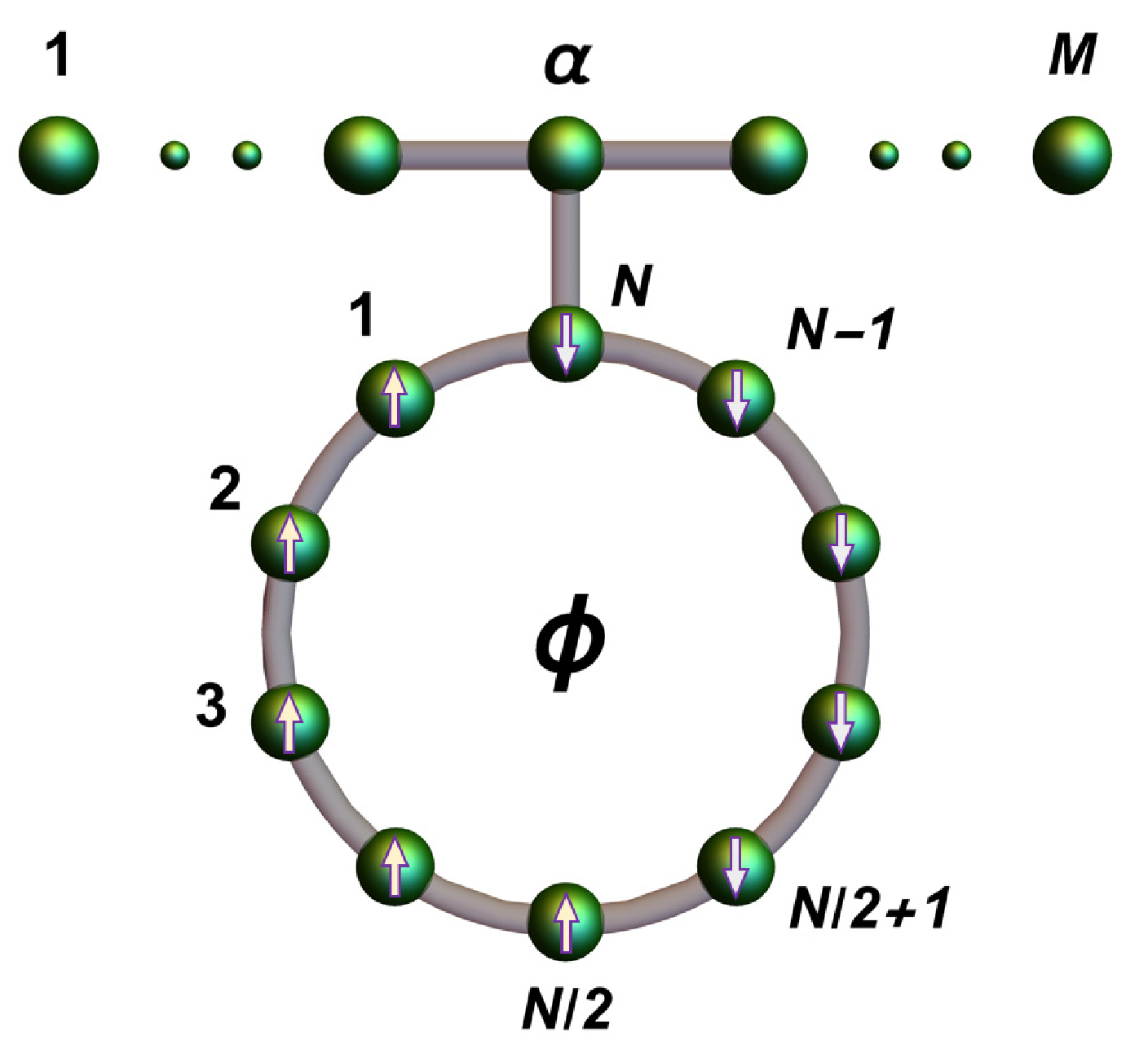}}\par}
\caption{(Color online). Schematic diagram of a ring-wire setup where an antiferromagnetic ring is coupled to a quantum wire. In one half
of the ring, magnetic moments are aligned along $+Z$ direction, while in the rest half the moments are oriented in the opposite direction,
to make the net magnetization zero. The ring encloses a magnetic flux $\phi$ which is responsible to produce a circular current in the ring.   
The side-attached quantum wire is assumed to be non-magnetic. The site number $N$ of the ring is coupled to the site number $\alpha$ of the 
wire.}
\label{model}
\end{figure}

In the presence of an AB flux, persistent charge current can be obtained in an isolated conducting loop and it is a well established
fact~\cite{van2000,matveev2002,majhi2021,majhi2023,cheung1988,cheung1989}. On the other hand, if the states are spin resolved, 
then it will be possible to obtain a net spin circular current in the antiferromagnetic ring. Moreover, if these charge and spin 
currents can be adjusted not only by the AB flux but also by other external factors, it could lead to new ways of controlling spin 
currents, which could be useful in designing advanced spintronic devices. The distinct characteristics of the antiferromagnetic 
ring, devoid of disorder, coupled with the Aubry-Andr\'e-Harper potential introduced in the chain (see Fig.~\ref{model}), provide a 
suitable example to unravel the interplay between ordered and disordered parts of a coupled system. 

Because of its unique and varied features, correlated disorder has received much more attention than completely 
uncorrelated (random) disorder.
Among various types, AAH disorder has been the subject of intense research in 
recent years~\cite{purkayastha2018,sahu2021,qi2023,mastropietro2015}. It is well-known that in a 1D system, the eigenstates become
exponentially localized when random disorder is introduced, leading to a critical disorder strength of zero. In contrast, for an AAH 
system, the transition from the delocalized to the localized phase occurs beyond a critical 
disorder strength~\cite{doggen2019,biddle2011,xing2022,sarkar2021}, which is a key feature of AAH disorder. Additionally, the AAH 
potential includes a phase factor, allowing the spectral properties to be tuned by adjusting this phase. These aspects make AAH 
disorder particularly noteworthy, and therefore, in our work we focus on AAH disorder.

In the present study, we investigate the interplay between AB effect, AF ring geometry, and AAH potential in mesoscopic rings. By employing 
a nearest-neighbor tight-binding model, we aim to explore and understand a wide range of intriguing phenomena. These phenomena include 
the spin channel splittings upon connecting the chain or a dangling bond to the antiferromagnetic ring, the energy level variations with
AB flux, AAH potential strength, AAH phase factor, and the dependence of charge and spin circular currents on these parameters as well 
as the system size. Our results will shed light on the underlying physics of these aspects and provide insights into the design and 
control of mesoscopic devices for future applications in quantum technologies.

The remainder of this study is divided into the following sections. In Sec. II, we present the quantum system and give an overview of 
the theoretical framework that is required to carry out the crucial computations. We use a TB framework to describe our system because 
it can provide a clear representation, especially when considering nanoscale dimensions. The results are obtained using well-known 
mathematical techniques that are well summarized for reference. In Sec. III, we present and critically analyze all the results. 
A brief description for realizing our proposed quantum system is given in Sec. IV. Finally, in Sec. IV we summarize 
our essential findings.

\section{Quantum system and theoretical framework}

\subsection{Description of the physical system and TB Hamiltonian}

The hybrid setup (see Fig.~\ref{model}) is constructed by  
two elements: an antiferromagnetic ring, threaded by an AB flux, and a one-dimensional wire subjected to an AAH potential. This 
combination provides a very good environment for studying different quantum phenomena at nanoscale level. The ring, 
consisting of $N$ lattice sites (where $N$ is even), is composed of two ferromagnetic semicircles of equal length, with opposite 
spin orientations along the $\pm Z$-axes (our chosen spin quantization directions). This configuration results in a net magnetization 
of zero. The chain, possessing $M$ lattice sites, is non-magnetic in nature and subjected to an AAH potential which plays the central
role of our study. The ring is directly connected to site $\alpha$ of the AAH chain through a single bond. For odd $M$, 
$\alpha$ is the central site of the chain, whereas for even $M$, $\alpha$ is just the next site of the central one. 
The inclusion of an AB flux in the ring imparts an additional phase factor, selectively affecting the ring and contributing to the 
emergence of persistent {\em charge} current. 

To describe the quantum system, we employ a tight-binding (TB) framework which is very convenient especially when the system is from 
electron-electron interaction.
We start with the total Hamiltonian $\mathcal{H}$ for our hybrid system under the nearest-neighbor hopping (NNH)
approximation~\cite{shokri2005,shokri2006,patra2022,bena2009}, which can be expressed as the sum of three components:
\begin{equation}
\boldsymbol{\mathcal{H}}=\boldsymbol{H}_{\text {Ring }}+\boldsymbol{H}_\text{C}+\boldsymbol{H}_{\text {Chain }.}
\end{equation}
Here, $\boldsymbol{H}_{\text{Ring}}$, $\boldsymbol{H}_\text{C}$, and $\boldsymbol{H}_{\text{Chain}}$ represent the TB Hamiltonians for 
the antiferromagnetic ring, the chain-ring coupling, and the chain, respectively. In our description, `wire' and `chain' have the same 
meaning. 

\vskip 0.2cm
\noindent
$\bullet $ \textbf{Hamiltonian of the antiferromagnetic ring ($\boldsymbol{H}_{\text{Ring}}$):}

\vskip 0.2cm
In our model, the ring is divided into two halves. In one half, the magnetic moments at different lattice sites 
(from the sites 1 to $N/2$) are oriented along the $+Z$ direction, while in the other half (from $N/2+1$ to $N$ sites), they are 
arranged along the $-Z$ direction. So, the effective magnetization of the ring is zero. The Hamiltonian $H_{\text{Ring}}$ is described 
as~\cite{shokri2005,shokri2006,patra2022}
\begin{eqnarray}
\boldsymbol{H}_{\text{Ring}} & =& \sum_n \boldsymbol{c}_n^{\dagger}\left(\boldsymbol{\varepsilon}_n^R - \boldsymbol{\vec{h}}_n \cdot \vec{\boldsymbol{\sigma}}\right) \boldsymbol{c}_n \\ \nonumber 
&& + \sum_n\left(\boldsymbol{c}_{n+1}^{\dagger} \boldsymbol{t} e^{i \Theta} \boldsymbol{c}_n + \text{h.c.}\right).
\end{eqnarray}
Here $\boldsymbol{c}_n^{\dagger}=\left(\begin{array}{cc}
c^{\dagger}_{n \uparrow} & c^{\dagger}_{n \downarrow}
\end{array}\right)$ and 
$ \boldsymbol{c}_n=\left(\begin{array}{c}
c_{n \uparrow} \\
c_{n \downarrow}
\end{array}\right)$. $c_{n\sigma}^{\dagger}$, $c_{n\sigma}$ are the conventional fermionic operators associated with site $n$ and 
spin $\sigma (\uparrow, \downarrow)$. The matrix $\boldsymbol{\varepsilon}_n^R=\left(\begin{array}{cc}
\varepsilon_{n \uparrow}^R & 0 \\
0 & \varepsilon_{n \downarrow}^R
\end{array}\right)$ involves on-site energies for up and down spin electrons in the absence of any magnetic interaction. 
The matrix $\boldsymbol{t}=\left(\begin{array}{cc}
t & 0 \\
0 & t
\end{array}\right)$ contains the nearest-neighbor hopping (NNH) strengths for two different spin cases. A phase factor $\Theta$ appears
in $\boldsymbol{H}_{\text{Ring}}$ due to the magnetic flux $\phi$ threaded by the ring, and it is expressed as 
$\Theta = 2 \pi \phi/N \phi_0$. The magnetic flux $\phi$ is measured in units of the elementary flux-quantum $\phi_0$~\cite{cheung1988}.
This flux is responsible to drive a current in the quantum ring. The interaction between an itinerant electron and the local magnetic
moment at any particular site $n$ is characterized by the term~\cite{sarkar2019,su2015,koley2021} $J\left\langle\overrightarrow{\boldsymbol{S}_n}\right\rangle \cdot \vec{\boldsymbol{\sigma}}$, 
where $\left\langle\overrightarrow{\boldsymbol{S}_n}\right\rangle$ denotes the average net spin at site $n$ and 
$\vec{\boldsymbol{\sigma}}$ is the Pauli spin vector. The parameter $J$ quantifies the strength of the spin-moment interaction. 
To simplify our notation and for convenience, we denote $J\left\langle\overrightarrow{\boldsymbol{S}_n}\right\rangle$ as 
$\vec{\boldsymbol{h}}_n$. This quantity is commonly known as the spin-dependent scattering parameter.

\vskip 0.2cm
\noindent
$\bullet $ \textbf{Hamiltonian of the coupling ($\boldsymbol{H}_\text{C}$):}

\vskip 0.2cm
The ring-wire coupling Hamiltonian can be written as
\begin{equation}
\boldsymbol{H}_{\mathrm{C}}=\left(\boldsymbol{c}_\alpha^{\dagger} \boldsymbol{\tau}\boldsymbol{c}_N +
\boldsymbol{c}_N^{\dagger}\boldsymbol{\tau} \boldsymbol{c}_\alpha\right)
\end{equation}
where we assume that the site number $N$ of the ring is coupled to the site number $\alpha$ of the wire (see Fig.~\ref{model}). 
$\boldsymbol{\tau}$ is a ($2\times 2$) diagonal matrix and it is written as $\boldsymbol{\tau}=\mbox{diag}(\tau,\tau)$, where $\tau$
measures the coupling strength between the wire and the ring.

\vskip 0.2cm
\noindent
$\bullet $ \textbf{Hamiltonian of the 1D chain ($\boldsymbol{H}_{\text{Chain}}$):}

\vskip 0.2cm
The Hamiltonian of the non-magnetic chain is expressed as
\begin{equation}
\boldsymbol{H}_{\text{Chain}} = \sum_n \boldsymbol{c}_n^{\dagger} \boldsymbol{\varepsilon}_n^C \boldsymbol{c}_n + 
\sum_n\left(\boldsymbol{c}_{n+1}^{\dagger} \boldsymbol{t} \boldsymbol{c}_n + \text{h.c.}\right)
\end{equation}
where $\boldsymbol{\varepsilon}_n^C=\mbox{diag}(\varepsilon_{n\uparrow}^C, \varepsilon_{n\downarrow}^C)$ and 
$\boldsymbol{t}=\mbox{diag}(t, t)$. $\varepsilon_{n\sigma}$'s are the site energies and $t$ is the NNH strength.
The chain is subjected to a correlated disorder in the form of Aubry-Andr\'e-Harper model. In presence of such a disorder, site energies
are expressed as $\varepsilon_{n\uparrow}^C=\varepsilon_{n\downarrow}^C=W \cos \left(2 \pi bn+\xi\right)$. In this expression three 
important parameters exist and they are as follows. The factor $W$ measures the cosine modulation strength. The other factor $b$ is an
irrational number and it creates the quasi-periodic modulation among the site energies. In our calculations, we set $b=(\sqrt{5}-1)/2$,
the most commonly used value in the literature, though any other irrational number can also be considered and the basic physics will be
undisturbed. The last parameter is $\xi$, which is refereed to as the AAH phase factor, and it can be tuned selectively with suitable
setup. The change of $\xi$ essentially modulates the site energies. Thus different arrangements of site energies can be obtained by
regulating the phase factor.
 
\subsection{Methodologies: Calculation of circular charge and spin currents}

In our investigation, we compute both charge and spin currents by employing the second-quantized formalism. Below we provide all the 
required important mathematical steps for a clear understanding.   

\subsubsection{Calculation of charge current}

We start with the charge current operator $\mathbf{I}_{\mathrm{c}}$. It is defined as~\cite{maiti2014,maiti2011}
\begin{equation}
\mathbf{I}_c=\frac{e \dot{\mathbf{x}}}{N a}
\end{equation}
where $e$ is the electronic charge, $\dot{\mathbf{x}}$ is the velocity operator, $N$ is the total number of lattice sites in the 
quantum ring and $a$ is the lattice spacing. 

The velocity operator can be expressed in terms of the position operator and the Hamiltonian as
\begin{equation}
\dot{\mathbf{x}}=\frac{1}{\mathrm{i} \hbar}[\mathbf{x}, \boldsymbol{\mathcal{H}}]
\end{equation}
where $\mathbf{x}$ represents the position operator and it is written in terms of the fermionic operators as
\begin{equation}
\mathbf{x}=\sum_n \mathbf{c}_n^{\dagger} n a \mathbf{c}_n.    
\end{equation}
Using these relations, and doing somewhat lengthy algebra we get the closed form of the charge current operator 
$\mathbf{I}_{\mathrm{c}}$ which looks like 
\begin{equation}
\begin{aligned}
\mathbf{I}_c= & \left(\frac{\mathrm{i} e t}{N \hbar}\right) \sum_m\left[\mathrm{e}^{-\mathrm{i} \Theta} c_{m \uparrow}^{\dagger} c_{m+1 \uparrow}-\text { h.c. }\right] \\
& +\left(\frac{\mathrm{i} e t}{N \hbar}\right) \sum_m\left[\mathrm{e}^{-\mathrm{i} \Theta} c_{m \downarrow}^{\dagger} 
c_{m+1 \downarrow}-\text { h.c. }\right] \\
= & \mathbf{I}_{\uparrow}+\mathbf{I}_{\downarrow}.
\end{aligned}
\end{equation}
Here, $\mathbf{I}_{\uparrow}$ and $\mathbf{I}_{\downarrow}$ signify the current operators for electrons with up and down spins, respectively.

To compute the current carried by an eigenstate $\left|\psi_k\right\rangle$, we employ the operation $\left\langle\psi_k\left|\mathbf{I}_{\mathrm{c}}\right| \psi_k\right\rangle$. This eigenstate $\left|\psi_k\right\rangle$ can be 
expressed as a linear combination of Wannier states, as follows
\begin{equation}
\left|\psi_k\right\rangle=\sum_p\left(a_{p \uparrow}^k|p \uparrow\rangle+a_{p \downarrow}^k|p \downarrow\rangle\right)
\end{equation}
where $a_{p \sigma}^k$'s are the coefficients. 
The current expression for the state $\left|\psi_k\right\rangle$ becomes
\begin{equation}
\begin{aligned}
I_c^k= & \left(\frac{\mathrm{i} e t}{N \hbar}\right) \sum_m\left[\mathrm{e}^{-\mathrm{i} \Theta} a_{m \uparrow}^{k \dagger} a_{m+1 \uparrow}^k-\text { h.c. }\right] \\
& +\left(\frac{\mathrm{i} e t}{N \hbar}\right) \sum_m\left[\mathrm{e}^{-\mathrm{i} \Theta} a_{m \downarrow}^{k \dagger} a_{m+1 \downarrow}^k-\text { h.c. }\right] \\
= & I_{\uparrow}^k+I_{\downarrow}^k.
\end{aligned}
\label{iuid}
\end{equation}
This equation reveals that the current carried by the eigenstate $\left|\psi_k\right\rangle$ can be written as the sum of 
up and down spin currents, denoted by $I_{\uparrow}^k$ and $I_{\downarrow}^k$ respectively.

To determine the net persistent charge current in the magnetic ring at absolute zero temperature, considering a specific chemical 
potential $\mu$, we sum over the lowest contributing energy levels. We express it as
\begin{equation}
I_{\mathrm{c}}=\sum_{k} I_{\mathrm{c}}^k.
\end{equation}
For the benefit of readers here we would like to point out that, some other methods to compute flux driven circular charge current 
are also available in the literature, and the simplest one is refereed to as the derivative method. But, compared to other existing 
techniques, the methodology described above is highly efficient as in this method we can evaluate the circular current in any particular 
loop of a multi-loop geometry. This facility is no longer available in other prescriptions.   

\subsubsection{Calculation of spin current}

Let us now look at the theoretical steps for computing another important component of our analysis which is the circular spin current.

Similar to the charge current operator, we define the spin current operator as~\cite{sun2008,sun2005,braunecker2007,davidson2020}
\begin{equation}
\mathbf{I}_{\mathrm{s}}=\frac{\overrightarrow{\mathbf{S}} \dot{\mathbf{x}}+\dot{\mathbf{x}} \overrightarrow{\mathbf{S}}}{2 a N}.
\end{equation}

In our investigation, we focus solely on the $Z$-component of the spin current, for the sake of simplification. Consequently, 
the spin current operator simplifies to
\begin{equation}
\mathbf{I}_{\mathrm{s}}^z=\frac{\hbar\left(\sigma_z \dot{\mathbf{x}}+\dot{\mathbf{x}} \sigma_z\right)}{4 a N}.
\end{equation}

By substituting $\sigma_z$ and $\dot{\mathbf{x}}$ into the equation and doing some algebra similar to charge current, we derive a closed 
form for the spin current operator. Using the operation $\left\langle\psi_k\left|\mathbf{I}{\mathrm{s}}^z\right| \psi_k\right\rangle$, 
we arrive at the expression for the spin current carried by an eigenstate $\left|\psi_k\right\rangle$ as
\begin{equation}
\begin{aligned}
I_{\mathrm{s}}^{z, k}= & \left(\frac{\mathrm{i} t}{2 N}\right)\left[\sum_m \mathrm{e}^{-\mathrm{i} \Theta} a_{m \uparrow}^{k \dagger} a_{m+1 \uparrow}^k-\text { h.c. }\right] \\
& -\left(\frac{\mathrm{i} t}{2 N}\right)\left[\sum_m \mathrm{e}^{-\mathrm{i} \Theta} a_{m \downarrow}^{k \dagger} a_{m+1 \downarrow}^k-\text { h.c. }\right] \\
= & \left(\frac{\hbar}{2 e}\right)\left(I_{\uparrow}^k-I_{\downarrow}^k\right).
\end{aligned}
\end{equation}
$I_{\uparrow}^k$ and $I_{\downarrow}^k$ are the current components associated with up and down spin electrons.

Finally, the net spin current at absolute zero temperature in the ring, for a particular chemical potential, is expressed as
\begin{equation}
I_{\mathrm{s}}^z=\sum_{k=1}^{N_{\mathrm{e}}} \boldsymbol{I}_{\mathrm{s}}^{z, k}.
\end{equation}

We can refer to $I_{\mathrm{s}}^z$ simply as $I_{\mathrm{s}}$, without any loss of generality, since we concentrate only on the 
$Z$-component of spin current.

\section{Numerical results and discussion}

In what follows we present and critically analyze our results one by one which essentially include the spectral behavior, circular
charge and spin currents under different input conditions like magnetic flux, AAH modulation, AAH phase factor, system size, ring-wire
coupling strength and other related things. 

Before delving into the discussion, let us mention the parameter values those are common throughout the numerical calculations. As 
already described earlier, the antiferromagnetic ring is made up with two ferromagnetic semicircles where in one semicircle the 
magnetic moments are aligned along $+Z$-direction, while for the other semicircle the moments are oriented along $-Z$. Thus, the polar
angles associated to the moments for these segments are $0$ and $\pi$, respectively. The strengths of the magnetic moments are assumed to
be identical viz, $h_n=h$ and set $h=1$. The site energies $\epsilon_{n \uparrow}$ and $\epsilon_{n \downarrow}$ in the ring are fixed
to zero, without loss of any generality. The NNH strengths $t$ and $\tau$ are set to $1$, unless specified. The AAH phase factor is also
considered to be zero, unless mentioned. All the energies are measured in units of electron-volt (eV). The rest other parameters those 
are not constant, are specified in the appropriate parts of the discussion.  
 
\subsection{Energy spectra}

To investigate the distinct characteristic features of circular charge and spin currents, it is indeed required to inspect the dependence
of energy eigenvalues under different input conditions associated with our chosen hybrid system. The first and foremost 
thing we need to 
check is whether there is a finite mismatch between the up and down spin energy levels. If no mismatch appears, we cannot expect any
spin-specific phenomenon. Additionally, the dependence of energy levels on the magnetic flux is crucial for generating a non-zero 
circular current in the ring.

\begin{figure}[ht]
{\centering \resizebox*{8cm}{6cm}{\includegraphics{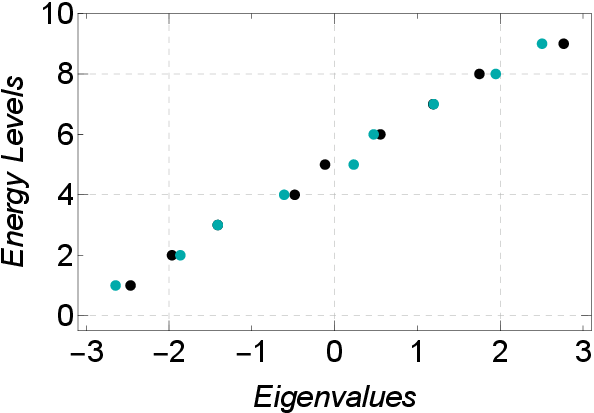}}\par}
\caption{(Color online). Splitting among the up (black dots) and down (cyan dots) spin energy levels for the wire-ring 
setup with $N=4$ and $M=5$. The magnetic flux $\phi$ is set at zero and AAH modulation strength $W$ is fixed at $1$. The mismatch among 
the energy eigenvalues associated with two different spin cases is clearly reflected.}
\label{esplit2}
\end{figure}

First, to check, whether we get a separation between up and down spin cases as we are dealing with a 
system possessing a zero net magnetization, in Fig.~\ref{esplit2} we plot the energy eigenvalues for up and down spin electrons 
considering a hybrid setup with $N=4$ and $M=5$. For this case, the magnetic flux is set at zero. 
Given that the magnetic moments are aligned along the $\pm Z$ directions, we can express the tight-binding Hamiltonian $\mathbf{H}$ 
as a sum of sub-Hamiltonians for up and down spin states, denoted as $\mathbf{H}_{\uparrow}$ and $\mathbf{H}_{\downarrow}$, 
respectively. Consequently, we have the opportunity to separately determine the energy eigenvalues associated with these sub-Hamiltonians.
In our chosen system, spin-dependent scattering occurs due to the existence of the factor $h$. In order to have a finite mismatch among 
the sets of up and down spin energy eigenvalues, we need to break the symmetry between up and down spin sub-Hamiltonians. If we consider
only the ring system (i.e., in the scenario where the ring-wire coupling is zero), the energy eigenvalues of two spin states exactly 
overlap to each other, as $\mathbf{H}_{\uparrow}$ and $\mathbf{H}_{\downarrow}$ are symmetric which can be understood by writing 
their TB Hamiltonian matrices. To break the symmetry in our setup, we couple the ring with a wire. While this wire can be an ordered 
one, we choose to use the AAH chain due to its intriguing properties instead of a simple ordered chain. Figure~\ref{esplit2} shows a 
mismatch among the energy eigenvalues associated with the up and down spin states. It gives a clear indication that spin specific phenomena
can be observed in our chosen hybrid setup.

\begin{figure}[ht]
{\centering \resizebox*{8.2cm}{5.5cm}{\includegraphics{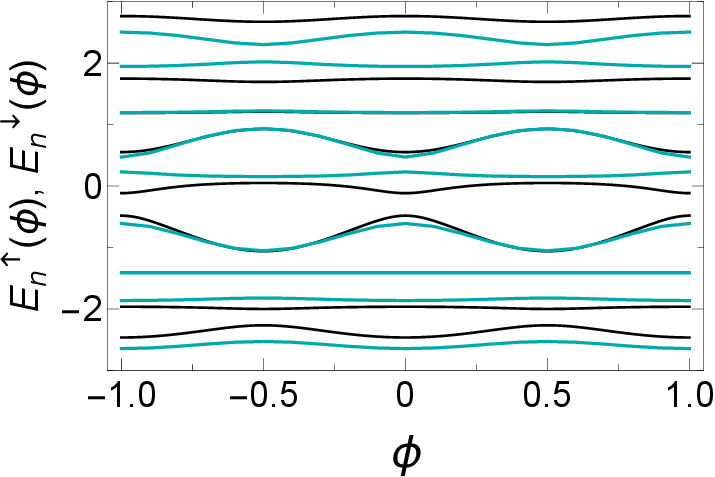}}\par}
\caption{(Color online). Variation of spin-dependent energy eigenvalues as a function of magnetic flux $\phi$ for the ring-wire coupled
system where the size of the system and the AAH modulation strength the same as mentioned in Fig.~\ref{esplit2}. The black and cyan colors 
are associated with up and down spin electrons respectively, like Fig.~\ref{esplit2}.}
\label{esplit3}
\end{figure}

Now we focus on the variation of spin-specific energy levels as a function of magnetic flux $\phi$. The results are shown in 
Fig.~\ref{esplit3} where we consider the identical system size and AAH modulation strength as used in Fig.~\ref{esplit2}. The response 
of the eigenenergies on the AB flux is highly important to achieve circular current in the ring.
Several interesting features are observed from Fig.~\ref{esplit3}. Following the symmetry breaking argument we can understand the mismatch
between up and down spin energy levels throughout the flux window. The most crucial thing is, in each spin case, we get a few number of 
energy levels that exhibit finite variations with flux $\phi$, while the rest other levels are independent of $\phi$. These later energy
levels (viz, those are flux-independent) do {\em not} carry any net current, as current carried by any state is directly linked with the
first order derivative of energy with respect to the magnetic flux. We refer these energy levels as the non-dispersive levels, and they are
associated with the quantum wire which is directly coupled to the ring system. The existence of zero and non-zero current carrying energy 
levels leads to several interesting aspects of circular current that can be understood from our forthcoming parts. In the absence of the 
wire, no such zero current carrying states appear.  

\begin{figure}[ht]
{\centering \resizebox*{8.2cm}{5.5cm}{\includegraphics{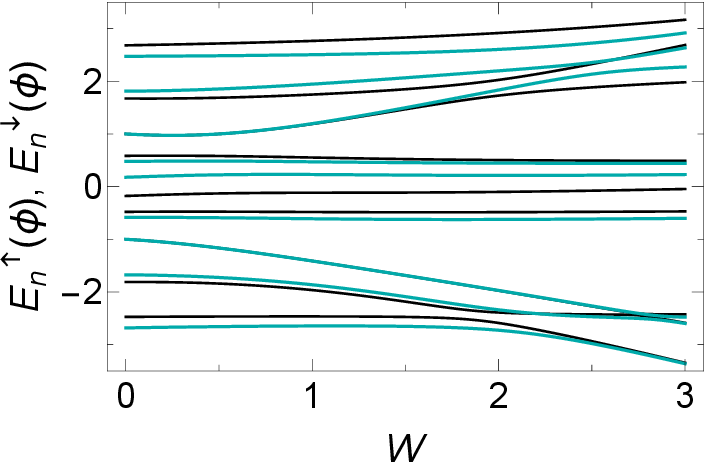}}\par}
\caption{(Color online). Dependence of up and down spin energy eigenvalues on AAH modulation strength $W$ for the wire-ring coupled 
system. The system size and the magnetic flux are the same as mentioned in Fig.~\ref{esplit2}. Two different colors correspond to the
identical meaning as prescribed in Fig.~\ref{esplit2}.}
\label{ew}
\end{figure}

In Fig.~\ref{ew}, we look at how energy levels change as we increase the AAH potential strength, with the AB flux set to zero. As we 
gradually increase the AAH potential, an interesting pattern appears: energy levels start to form clear sub-bands, which are closely 
tied to the strength of the AAH potential. Notably, energy levels between $-1 \text{eV}$ and $+1 \text{eV}$ show less sensitivity to 
changes in the AAH potential. These levels remain relatively stable even as the disorder strength increases, indicating a certain 
robustness in this energy range. This stability is likely because these energy states are less localized and more spread out, even 
in the presence of disorder.
On the other hand, energy levels outside this range are much more sensitive to changes in the AAH potential, with their energy range
significantly widening as disorder increases. The AAH disorder can cause a transition between delocalization and localization. At lower 
disorder strengths, electrons can move freely through the chain, forming extended states. But as the disorder strength increases, 
quantum interference effect becomes more pronounced, making certain regions of the chain energetically unfavorable for electron 
movement due to destructive interference. This results in a decrease in the movement of electrons through the wire, causing them to 
primarily travel within the ring. We can describe our hybrid system as an ordered-disordered separated (ODS) system, 
where an `ordered' segment is connected to a `disordered' one. The ODS systems are particularly intriguing when compared to both fully 
ordered and fully disordered systems. In the regime of weak disorder, the ordered segment becomes increasingly disturbed by the 
disordered region, with this disturbance growing as $W$ increases. The coupling between the two regions, however, is 
inversely proportional to $W$. In the case of strong disorder, the ordered and disordered segments become almost decoupled, and 
under these conditions, only the ordered region (in our case, the ring system) contributes to the current.  

\subsection{Charge Current}

\begin{figure}[ht]
{\centering \resizebox*{8.2cm}{5.5cm}{\includegraphics{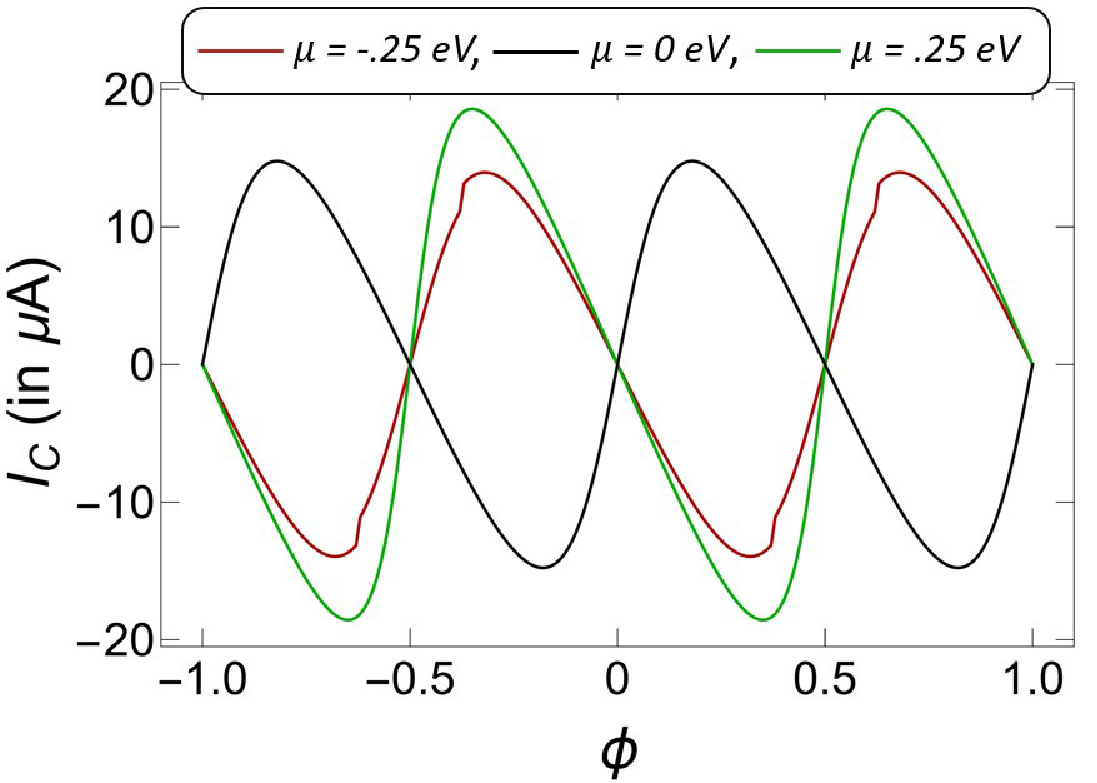}}\par}
\caption{(Color online). Variation of circular charge current as a function of AB flux $\phi$ for the wire-ring setup. Here we select $20$
sites in the ring and $50$ sites in the side-coupled chain. Three different colored curves are associated with three distinct values of 
the chemical potential $\mu$. Here $W=1$.}
\label{iph}
\end{figure}

In Fig.~\ref{iph}, we examine the behavior of circular charge currents in response to variations in AB flux, $\phi$, within our composite
system, which consists of a magnetic ring with $20$ lattice sites and a non-magnetic AAH chain with $50$ sites. We calculate the 
circular charge currents using Eq.~\ref{iuid} across a wide range of AB flux values, highlighting their behavior under three distinct 
chemical potentials ($\mu$) set at $-0.25$, $0$, and $0.25$. The results are presented in three distinct colored curves, each corresponding
to a different $\mu$. 
A detailed analysis reveals that the charge current's behavior is distinct as the AB flux is varied, and this behavior is sensitively 
influenced by the chemical potential $\mu$. One of the most notable observations is the periodicity of the charge currents with respect 
to the AB flux, indicating the system's sensitivity to changes in external magnetic flux. These circular charge currents exhibit periodic
oscillations across all considered values of the chemical potential. However, the current profiles show intriguing differences depending 
on the value of $\mu$. 
Specifically, when $\mu$ is set to zero, the polarity of the charge current reverses compared to the other two 
cases. This significant shift in current direction highlights the profound impact of chemical potential on charge transport within the 
system. Additionally, the magnitudes of the circular charge currents vary notably with changes in $\mu$, revealing the sensitivity of 
charge transport to the energy levels defined by the chemical potential.

\begin{figure}[ht]
{\centering \resizebox*{8.2cm}{5.5cm}{\includegraphics{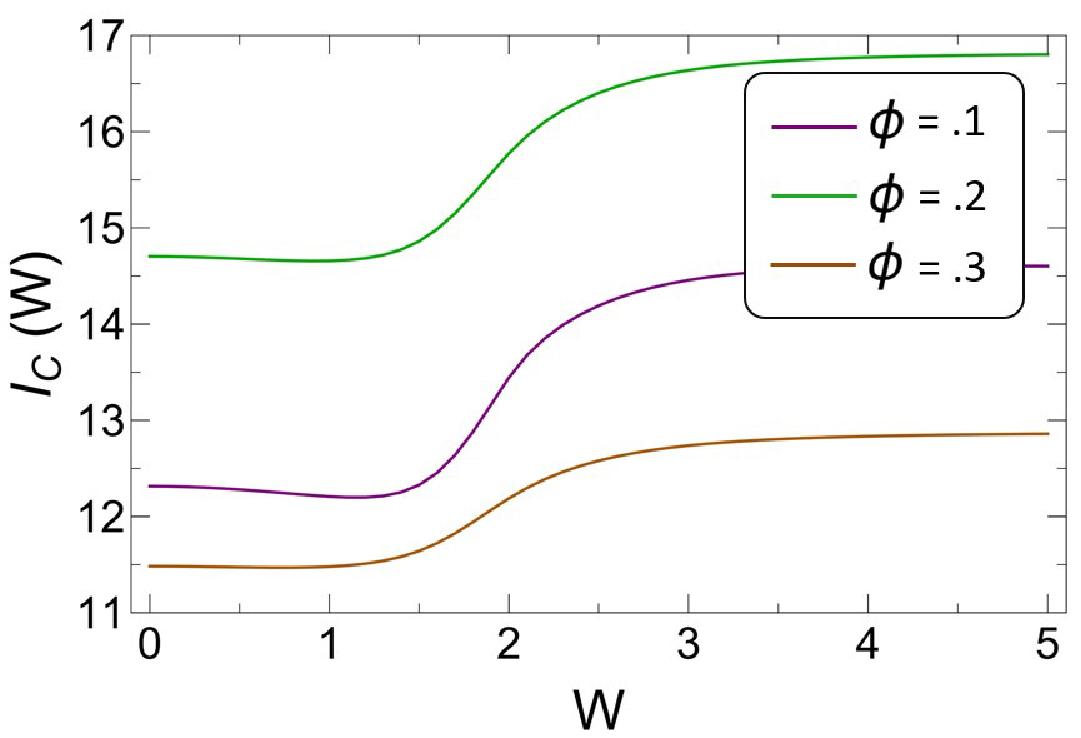}}\par}
\caption{(Color online). Persistent charge current as a function of AAH modulation strength for three different values of AB flux $\phi$, 
where the purple, green, and orange curves correspond to $\phi=0.1$, $0.2$, and $0.3$, respectively. Here we set $\mu=0$. The size of
the ring-wire setup remains the same as used in Fig.~\ref{iph}.}
\label{iw}
\end{figure}

Figure~\ref{iw} provides a detailed view of how the charge current evolves with varying AAH potential strength $W$.
The currents are computed at three typical values of $\phi$ by changing the AAH modulation strength in a wide range, for the identical 
system size as taken in Fig.~\ref{iph}.  
When we closely examine the charge current plots, a clear pattern emerges. Initially, as the AAH potential strength increases, the 
charge current stays relatively stable, almost constant. This suggests that the AAH potential has a weak effect on charge transport. 
However, when the AAH potential strength crosses a certain threshold, things change significantly. The charge current starts to increase,
indicating that stronger AAH potential enhances charge transport. This increase in current eventually levels off, meaning it stops growing. 
This saturation highlights the complex interaction between disorder-induced localization and extended electron states. 
It is important 
to note that disorder is only introduced in the chain while the ring remains free of disorder, creating a situation similar to a separation
between ordered and disordered regions. As already discussed earlier, at higher AAH potential strengths, the chain effectively decouples 
from the ring, leaving the electronic states of the ring unaffected by the chain. As a result, the ring exhibits conducting states, leading 
to increased circular current at higher $W$. In our wire-ring coupled system, we can describe this as `disorder-induced delocalization', 
which could be useful in studying transport behavior in other contexts. When $W$ is extremely large and the ring is completely decoupled 
from the disordered chain, the current reaches a saturation point.

\begin{figure}[ht]
{\centering \resizebox*{8.2cm}{5.5cm}{\includegraphics{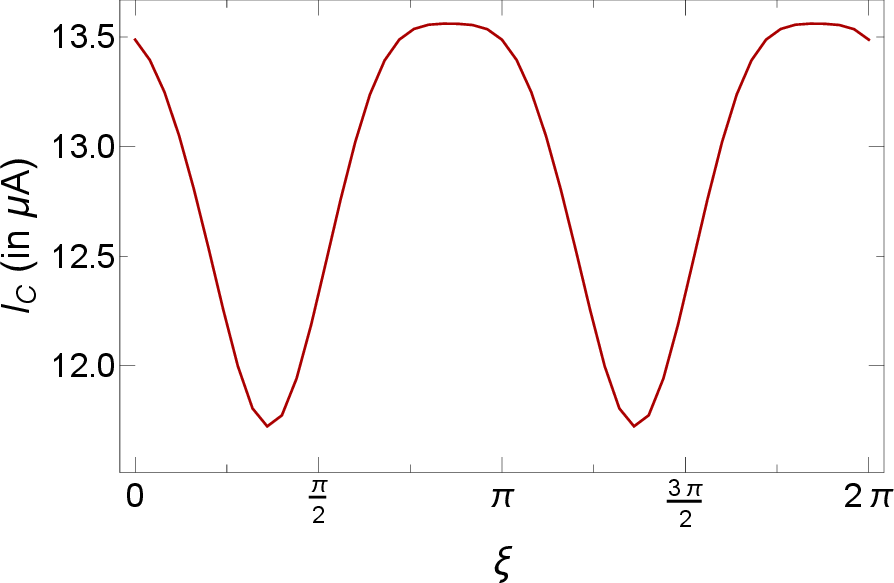}}\par}
\caption{(Color online). Dependence of circular charge current on the AAH phase factor for the wire-ring setup. The system size remains
the same as taken in Fig.~\ref{iph}. Here we set $\mu=0$, $\phi=0.25$, and $W=1$.}
\label{ixi}
\end{figure}

The variation of the charge current with the AAH phase factor $\xi$ is depicted in Fig.~\ref{ixi} for the same system size described 
earlier. We have uncovered intriguing oscillatory behavior in the charge current as a function of the AAH phase. This phenomenon arises 
from the interplay between the AAH potential, geometry of the system, and quantum interference effects.
As the AAH phase is modified, the configuration of the AAH potential along the chain is effectively altered. This induces a finite 
variation in the onsite energies of the atomic sites within the chain, leading to changes in the energy levels of the electronic 
structure. As the AAH phase varies, the energy levels shift correspondingly, impacting the ability of electrons to propagate through 
the chain. When the energy levels align favorably with the chemical potential of the system, electrons can traverse the chain more 
easily, and electronic wave functions interfere constructively, resulting in enhanced charge currents. Conversely, when the energy 
levels deviate from the chemical potential, electron transport becomes less efficient, leading to reduced charge currents.

\subsection{Spin Current}

Now, in this sub-section, we discuss the phenomenon of circular spin current.
\begin{figure}[ht]
{\centering \resizebox*{8.2cm}{5.5cm}{\includegraphics{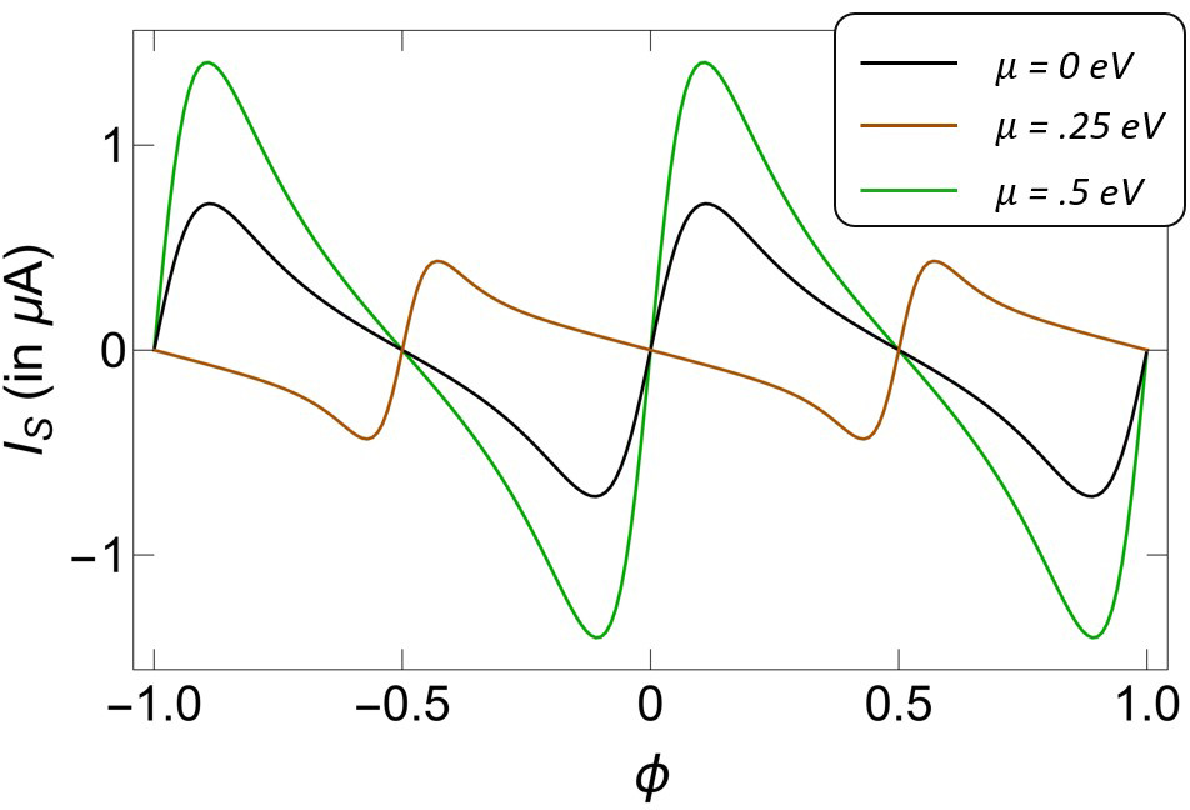}}\par}
\caption{(Color online). Circular spin current as a function of AB flux $\phi$ for the ring-chain setup, where the ring and the chain 
contain $20$ and $50$ lattices, respectively. The currents are calculated at three typical values of $\mu$, where the black, orange
and the green curves correspond to $\mu=0$, $0.25$, and $0.5$, respectively. The AAH disorder strength in the chain is fixed at $1\,$eV.}
\label{isph}
\end{figure}
\begin{figure}[ht]
{\centering \resizebox*{8.2cm}{5.5cm}{\includegraphics{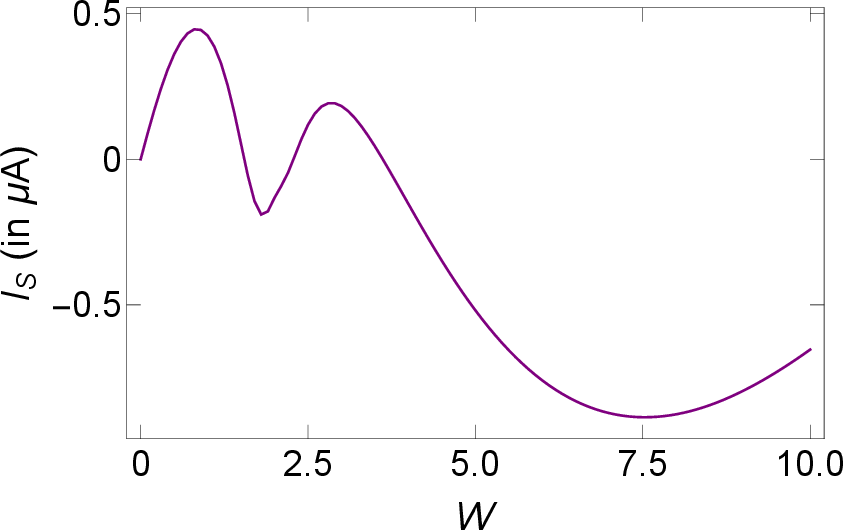}}\par}
\caption{(Color online). Variation of spin current as a function of AAH modulation strength $W$ at a constant AB flux ($\phi = 0.25$) 
for the wire-ring system. Here, $\mu$ is set at zero.}
\label{isw}
\end{figure}
In Fig.~\ref{isph} we show the characteristic features of spin current as a function of AB flux $\phi$ for the composite system. 
Three different values of $\mu$ are taken into account and in each case we set $W=1\,$eV.
A closer examination of the spin current plots reveals an intriguing behavior: the spin current exhibits periodic oscillations as the 
AB flux varies. This periodicity stems from the system's sensitivity to changes in the external magnetic flux, which modulates the spin 
currents as the AB flux is adjusted, resulting in the observed oscillatory patterns. These oscillations are a direct consequence of 
electron wavefunction interference within the composite system, leading to constructive and destructive interference patterns that 
cause periodic variations in the magnitude and direction of the spin current.

Further, a distinctive observation arises when comparing the spin current profiles at $\mu=0.25\,$eV with those at $\mu=0$ and 
$\mu=0.5\,$eV. The spin current for $\mu=0.25\,$eV exhibits an opposite polarity in comparison to the other two cases. This reversal in 
current direction highlights the significant impact of chemical potential on spin transport within the system. The energy landscape created 
by $\mu=0.25\,$eV results in a distinct behavior of electron states, leading to the observed change in current polarity.
Moreover, the magnitudes of the spin current differ for each chemical potential value. This variation in current magnitude can be attributed 
to the specific energy states available within the system at different chemical potentials. The chemical potential essentially determines 
the availability of energy states for electrons to occupy, and the resulting population of these states affects the overall spin current. 
Thus, the differences in spin current magnitudes for various chemical potentials are a reflection of the varying electron population and 
their associated spin properties within the composite system.

In Fig.~\ref{isw}, we explore how the spin current changes as the AAH potential strength $W$ varies, while keeping the AB flux constant 
at $\phi = 0.25$. The system size is the same as in the previous discussion, and the chemical potential is set to zero. 
\begin{figure}[ht]
{\centering \resizebox*{8.2cm}{5.5cm}{\includegraphics{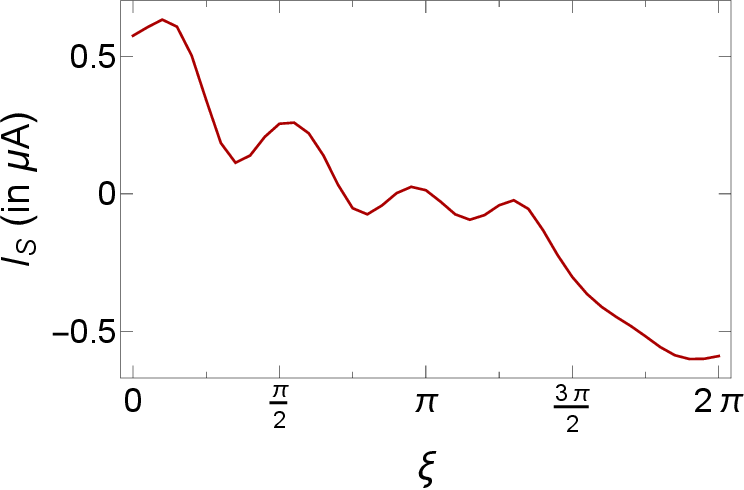}}\par}
\caption{(Color online). Dependence of spin current on the AAH phase factor $\xi$ for the ring-wire coupled system. Here $\mu=0$, 
$\phi=0.25$, and $W=1$.}
\label{isxi}
\end{figure}
Our findings reveal a fascinating and complex pattern in the spin current as $W$ is adjusted. Instead of a simple periodic oscillation, 
the spin current alternates between positive and negative values, reflecting how the system reacts to changes in the AAH potential strength.
This behavior is due to the interaction between the AAH potential and the system's geometry. The AAH potential creates irregular changes 
in the onsite energies of the chain, which affects the energy levels of the electronic states. This modulation can also lead to the 
formation of localized electronic states that alternate between positive and negative characteristics. As a result, the spin current 
shifts between positive and negative as the AAH potential strength changes, demonstrating the system's sensitivity to the unique effects 
of the AAH potential.

In Fig.~\ref{isxi}, we show how the spin current changes with the AAH phase factor $\xi$. Unlike what has been seen before, the spin 
current here does not follow a regular pattern. Instead, it shows quite an irregular oscillation as the AAH phase factor changes.
At first, when the AAH phase factor starts from zero and increases to $\xi=2\pi$, the spin current stays positive for some values of 
the phase, but then it turns negative. This shift happens because the phase factor causes significant changes in the electronic 
states, leading to a reversal in the direction of the spin current. As $\xi$ continues to increase, the spin current becomes negative,
reflecting changes in the electronic configurations that now favor spin-polarized electrons moving in the opposite direction.
Throughout this phase variation, there are also small peaks of opposite polarity in the spin current. At these phases, certain electronic 
states become particularly favorable for spin transport, leading to brief spikes in the spin current.

\subsection{Effect of System Size}

Since the energy spectrum gets significantly influenced by the side energy modulation in the quantum wire coupled to the ring system, 
\begin{figure}[ht]
{\centering \resizebox*{8.2cm}{11.cm}{\includegraphics{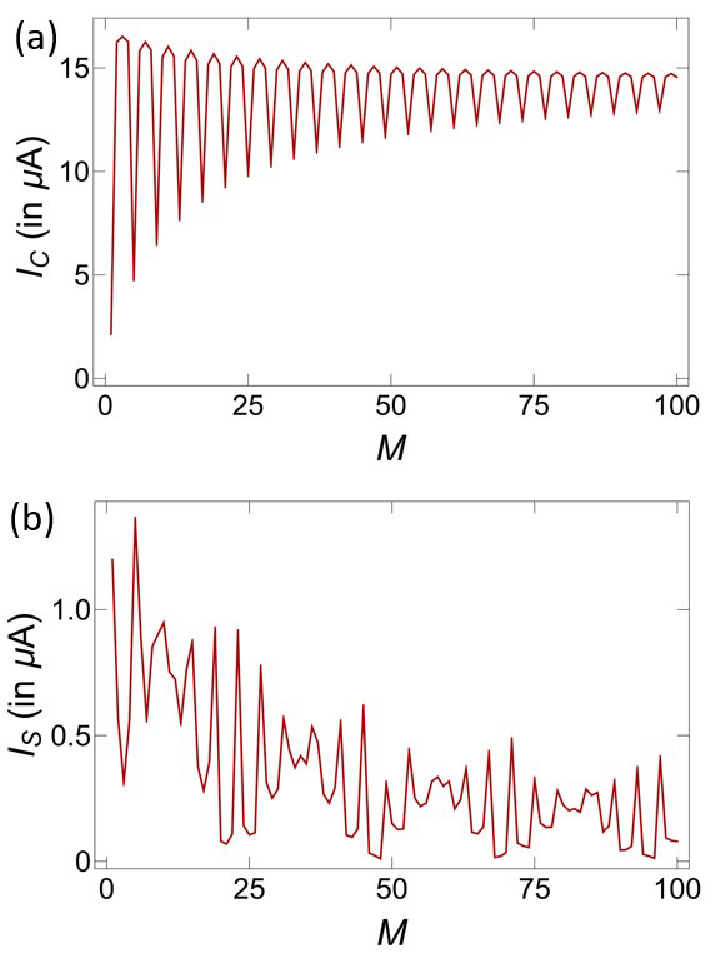}}\par}
\caption{(Color online). Charge and spin circular currents, shown in (a) and (b) respectively, as a function of AAH chain length (viz,
$M$) when the ring size $N$ is kept constant. We select $N=20$ and choose $\phi=0.25$, $\mu=0$, and $W=1$.}
\label{iM}
\end{figure}
it is quite natural to inspect the dependence of chain length (viz, $M$) on circular charge and spin currents. The results are shown in
Fig.~\ref{iM} where (a) and (b) are associated with the charge and spin currents respectively. Here we fix the ring size $N$ and vary
the chain length in a wide range.  

When we look at how the charge circular current behaves, we notice an interesting pattern. At first, with a small chain size, the current 
drops quickly and reaches its lowest point. But then, the current starts to recover and gradually increases. This recovery happens because 
a larger Aubry chain provides more pathways for electrons to move through, making it easier for them to flow and increasing the current.
On the other hand, the maximum values of the current start to decrease. This might be because, in larger chains, even though there are 
more pathways, certain patterns or effects slow down the current.
As the size of the Aubry chain increases, we see rapid changes in the current due to interference caused by disorder. 
In smaller chains, 
these interference effects are stronger because there are fewer atomic sites. This causes the current to fluctuate quickly. However, as 
the chain gets bigger, the interference effects lessen, leading to a smoother recovery and eventually an increase in the current.

The variation in the spin circular current with increasing chain size reveals an interesting pattern that sheds light on the basic 
physics of charge transport in mesoscopic systems subjected to disorder. Initially, when the chain size is small, typically in the range 
of $1$ to $10$ atomic sites, the spin circular current exhibits relatively high values. However, as the chain length continues to increase, 
the behavior of the spin circular current becomes more intriguing. It starts to decrease and undergo oscillations. This phenomenon can be
attributed to the increasing complexity of the system as the chain size grows. Disorder-induced localization effects become more pronounced,
leading to a reduction of the spin circular current amplitude. Furthermore, the oscillatory nature of the spin circular current arises 
from the interference phenomena. In larger chains, there are more opportunities for electrons to scatter and interfere with each other,
resulting in oscillations of spin circular current. Despite the oscillation and localization effects, the spin circular current does not 
exhibit a continuous reduction. It demonstrates moments of recovery, indicating that specific chain sizes permit the emergence of extended
electron states that can positively contribute to the spin circular current. 

\subsection{Effect of Coupling}

\begin{figure}[ht]
{\centering \resizebox*{8.2cm}{11 cm}{\includegraphics{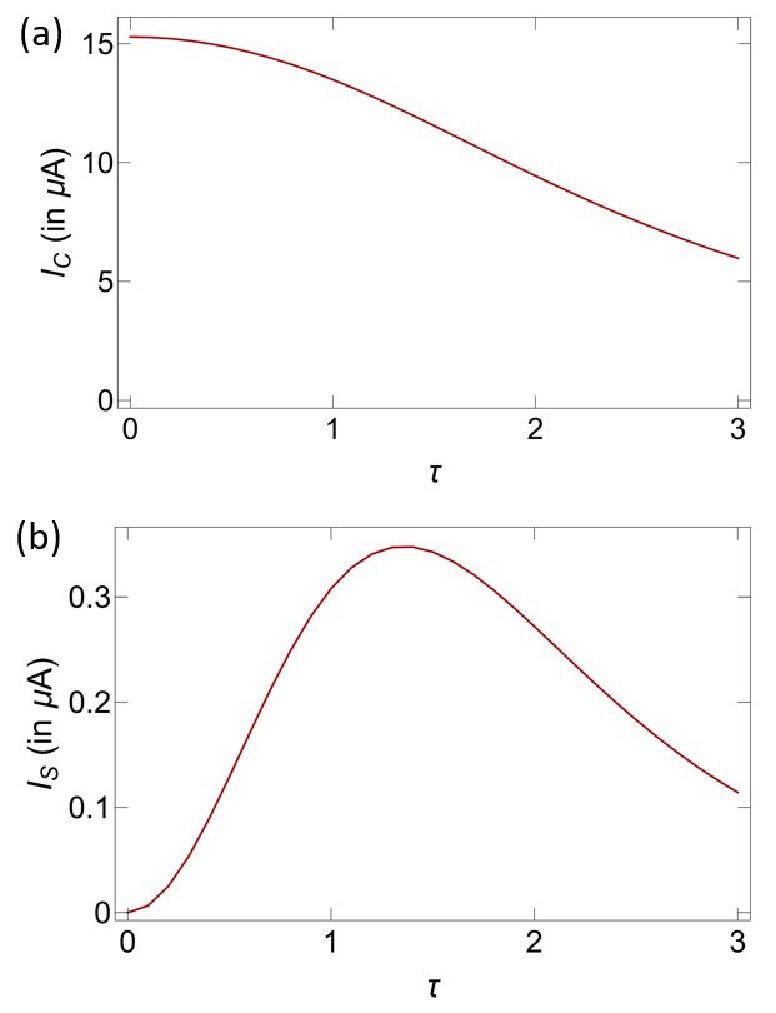}}\par}
\caption{(Color online). Dependence of ring-to-chain coupling on circular charge and spin currents, shown in (a) and (b), respectively.
Like the cases, here also we choose $20$ sites in the ring and $50$ sites in the chain. The other parameters are: $\mu=0$, $\phi=0.25$,
and $W=1$.}
\label{iC}
\end{figure}

Now, we explore the specific role of ring-to-chain coupling on circular charge and spin currents. The results are shown in 
Fig.~\ref{iC}, where 
(a) and (b) correspond to the charge and spin circular currents respectively. Several interesting features are obtained those are as follows. 
At first glance, we observe that the charge current gradually decreases as the ring-wire coupling strength $\tau$ increases. The circular 
charge current for a given filling mainly depends on the contribution from the available energy channels. As previously 
discussed, due to 
the presence of the AAH chain, non-dispersive energy channels appear in the band spectrum, along with current-carrying states originating 
from the ring. Since the wire is directly connected to the ring, the ring states are also influenced by the chain. For a specific $W$,
increasing $\tau$ causes more disruption to the ring states, leading to greater localization. As a result, the overall charge current 
decreases with $\tau$.

On the other hand, the behavior of the spin current is different and more intriguing compared to the charge current. For a spin current 
to exist, there must be a break in symmetry between the up and down spin sub-Hamiltonians. In our system, this is achieved by adding a 
wire to the anti-ferromagnetic ring. When the coupling strength $\tau$ is zero, no spin current is generated. However, when $\tau$ is set 
to a non-zero value, the symmetry is broken, leading to a finite spin current. As the coupling strength increases, more asymmetry is 
introduced, which initially enhances the spin current. However, because the chain is disordered, increasing the coupling strength also 
disturbs the ordered part of the system (the ring), reducing both the up and down spin currents. Consequently, the overall spin current
decreases with stronger wire-ring coupling. This is particularly interesting because the spin current can be selectively tuned by 
simply adjusting the coupling strength $\tau$.

\section{Possible experimental realization}

For the sake of completeness of the study, we finally discuss about the possible experimental realizations of our 
chosen quantum system. Our system consists of an antiferromagnetic ring and a quantum wire, where the later one is subjected to 
cosine modulation in the form of AAH model. By placing these two components, viz the ring and the wire, in close proximity, a direct
coupling between them can easily be established, and similar kind of configurations with non-magnetic and magnetic systems have also 
been used for studying other aspects. The designing of our chosen one-dimensional antiferromagnetic is very simple as it is made up of
two ferromagnetic chains of identical length. In fact, nowadays, systems with different kinds of magnetic arrangements can be fabricated 
quite easily with the use of sophisticated fabrication technologies\cite{winzer1996,weekes2004,jariwala2001,cui2002,tong2008,an2005}. 
On the other hand, an AAH chain can 
be constructed with the help of two counter propagating laser beams have waving different vectors, say, $K_1$ and $K_2$. The ratio of
these wave vectors measures the incommensuration factor of the potential. Creating the profile, the atoms are trapped in the dip regions.
This is the standard prescription and has been well established in the literature\cite{amico2005,frankearnold2007,aghamalyan2013}. Changing anyone of the two
among $K_1$ and $K_2$ or the both, we can have different distribution of the cosine modulation, which is equivalent to the change of
the AAH phase factor. Thus, we can safely claim that, our proposed ring-wire hybrid system can be designed and the results studied here
can be examined in a suitable laboratory.

\section{Closing Remarks}

In this work, we studied charge and spin circular currents in a hybrid system where an antiferromagnetic AB ring is directly connected 
to a non-magnetic chain. We introduced correlated disorder in the chain using the AAH model to explore how the interaction 
between 
ordered and disordered regions affects the circular currents. Since correlated disorder often leads to more intriguing phenomena 
compared to uncorrelated (random) disorder, we chose the AAH model. A tight-binding framework was used to describe the quantum system, 
and all results were derived using standard theoretical methods. The findings were carefully analyzed under various physical conditions, 
leading to the following key insights. \\
$\bullet$ We have observed interesting periodic changes in charge currents when exposed to AB flux, showing the system's sensitivity 
to external magnetic flux. The behavior of these currents varies noticeably with $\mu$, even causing the current to reverse direction, 
which highlights the important role of chemical potential in charge transport.\\
$\bullet$ When disorder is present only in the chain, it creates a mix of ordered and disordered regions. As the disorder strength 
increases, the connection between the ring and the chain weakens, causing the particles moving through the ring to be less disturbed. 
This leads to an increase in current with $W$.\\
$\bullet$ Changes in the AAH phase cause shifts in energy levels, which impact how electrons move through the chain. When these energy 
levels align with the chemical potential, charge currents increase. However, when they don't align, electron transport is disrupted, 
leading to reduced currents. \\
$\bullet$ Adding the AAH chain disrupts the symmetry between up and down spin channels, creating spin currents. This shows how important 
the connection between the ring and the chain is in this process.\\
$\bullet$ The direction of the spin current changes with the chemical potential and oscillates regularly with changes in AB flux. 
This back-and-forth behavior of charge and spin currents is due to interference effects caused by disorder-induced localization. \\
$\bullet$ When there is no coupling between the antiferromagnetic ring and the AAH chain, the spin current is zero because the system 
is symmetric. As the coupling strength increases, spin current starts to rise because the AAH chain breaks this symmetry. However, 
after a certain point, further increasing the coupling strength causes destructive interference, which reduces the spin current. 
This shows the complex relationship between coupling, symmetry breaking, and charge and spin transport in small-scale systems.

Our study offers valuable insights into how symmetry, disorder, and electronic transport interact in small-scale systems. These findings 
help us better understand quantum transport and could be useful for designing and improving spintronic and quantum devices by taking 
advantage of symmetry-breaking effects.\\

\textbf{Data Availability Statement:} The data that support the findings of this study are available upon reasonable request from the authors.

\bibliographystyle{apsrev4-1}

\bibliography{bibtex.bib}

\end{document}